\documentclass{ieeeaccess}
\usepackage{cite}
\usepackage{amsmath,amssymb,amsfonts,amscd}
\usepackage{braket}
\newtheorem{thm}{Theorem}[section]

\newtheorem{cor}{Corollary}

\newtheorem{exmp}{Example}[section]

\usepackage{arydshln}

\usepackage{algorithm}
\usepackage{algpseudocode}
\usepackage{graphicx}
\usepackage{caption}
\usepackage{tabstackengine}
\usepackage{hyperref}

\usepackage{textcomp}

\def\BibTeX{{\rm B\kern-.05em{\sc i\kern-.025em b}\kern-.08em
    T\kern-.1667em\lower.7ex\hbox{E}\kern-.125emX}}

\usepackage[thinlines]{easybmat}
\usepackage{multirow,bigdelim}

\begin{document}
\doi{10.1109/TQE.2020.DOI}
\title{Engineering Quantum Error Correction Codes Using Evolutionary Algorithms}
\author{\uppercase{Mark A. Webster}\authorrefmark{1}
and \uppercase{Dan E. Browne}.\authorrefmark{1}}
\address[1]{Department of Physics \& Astronomy, UCL,
Gower Place, London, WC1E 6BN}
\tfootnote{``This work was supported in part by the Engineering and Physical Sciences Research Council on Robust and Reliable Quantum Computing (RoaRQ), Investigation 011 [grant reference EP/W032635/1] and by the Engineering and Physical Sciences Research Council [grant number EP/S005021/1]. For the purpose of open access, the author(s) has applied a Creative Commons Attribution (CC BY) license to any Accepted Manuscript version arising.''}

\markboth
{Author \headeretal: Preparation of Papers for IEEE Transactions on Quantum Engineering}
{Author \headeretal: Preparation of Papers for IEEE Transactions on Quantum Engineering}

\corresp{Corresponding author: Mark A. Webster (email: mark.acacia@gmail.com).}

\begin{abstract}
Quantum error correction and the use of quantum error correction codes is likely to be essential for the realisation of practical quantum computing.
Because the error models of quantum devices vary widely, quantum codes which are tailored for a particular error model may have much better performance.
In this work, we present a novel evolutionary algorithm which searches for an optimal stabiliser code for a given error model, number of physical qubits and number of encoded qubits.
We demonstrate an efficient representation of stabiliser codes as binary strings - this allows for random generation of valid stabiliser codes, as well as mutation and crossing of codes.
Our algorithm finds stabiliser codes whose distance closely matches the best-known-distance codes of \cite{codetables.de} for  $n \le 20$ physical qubits.
We perform a search for optimal distance CSS codes, and compare their distance to the best-known-codes. 
Finally, we show that the algorithm can be used to optimise stabiliser codes for biased error models, demonstrating a significant improvement in the undetectable error rate for $[[12,1]]$ codes versus the best-known-distance code with the same parameters.
As part of this work, we also introduce an evolutionary algorithm QDistEvol for finding the distance of quantum error correction codes.
\end{abstract}


\titlepgskip=-15pt

\maketitle

\section{Introduction}
\label{sec:introduction}
\PARstart{Q}{uantum} computers have the potential to solve problems beyond the 
 capabilities of classical computers. 
One of the main challenges in reaching this potential is protecting the information used in quantum computers from errors due to environmental interactions and imperfectly executed operations.
Quantum error correction is one of the proposed methods to address this challenge, and involves the use of quantum error correction codes \cite{CSS} which can detect and correct errors during calculations.
Stabiliser codes \cite{stabiliser_codes} are one of the most commonly used types of quantum error correction code. 
Quantum devices employ a wide range of qubit types and so errors may arise quite differently on each device.
As a result, a stabiliser code which works well for one device may not work well on another.

A number of different works have examined tailoring stabiliser codes for use in devices with different error models\cite{optimisation_decoder,optimisation_RL,optimisation_QC,optimisation_cartan, optimisation_iterative,optimisation_convex,kubica_surface}. 
As the search space for stabiliser codes grows exponentially with the number of physical and logical qubits, these works generally focus on stabiliser codes on a small number of qubits, or search for variations of a particular stabiliser code . 

\begin{figure}[h]
    \centering
    \includegraphics[width=0.95\linewidth]{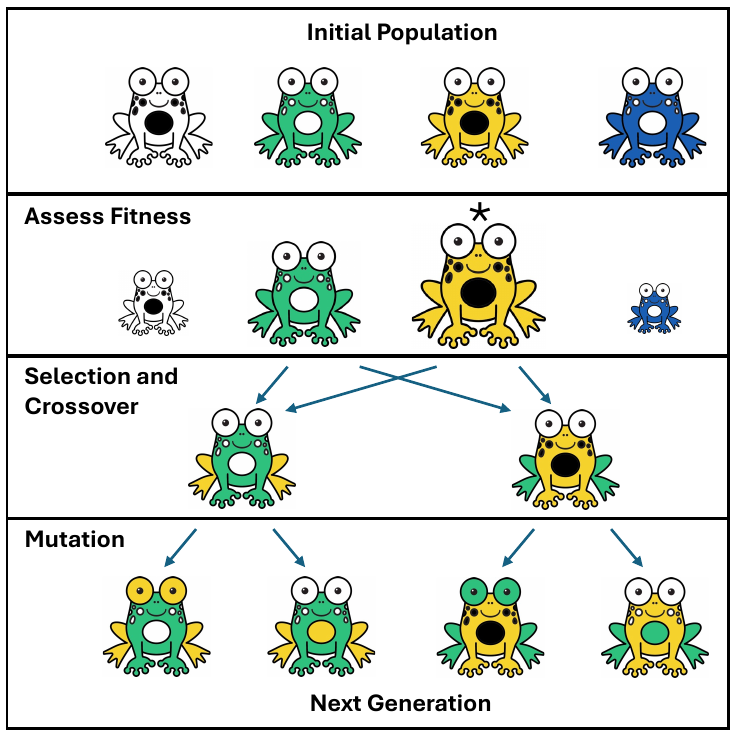}
    \caption{Outline of the evolutionary algorithm. We start with a random initial population of size $\lambda$. We then assess the fitness of each individual and  keep track of the individual with the best fitness score so far (starred yellow frog). We then select the $\mu$ individuals with the best fitness to form the reproducing pool.
    Individuals in the reproducing pool are crossed to form $\mu$ new individuals. The next generation is  formed by creating $\lambda/\mu$ mutations of each of the crossed individuals.
    This process is repeated for a specified maximum number of generations, and the individual with the highest observed fitness is returned.}
    \label{fig:evolutionary_diag}
\end{figure}

In this work, we develop an evolutionary algorithm to find a globally optimal stabiliser code for a given error model in terms of the rate of undetectable errors.
Evolutionary algorithms can find global optima in large search spaces and involve optimisation over a number of generations. The main features of evolutionary algorithms are set out in Fig.~\ref{fig:evolutionary_diag}.
We show how to encode  stabiliser codes into a binary string which serves as a genotype for the evolutionary algorithm. 
Each string of a particular length represents a valid stabiliser code, allowing random stabiliser codes to be generated as well as crossing and mutating individuals at each generation. 
Calculating the undetectable error rate exactly is computationally expensive for stabiliser codes with a large number of physical and logical qubits and we demonstrate a method for approximating this.

We benchmark our algorithm by searching for high distance codes for a given number of physical and logical qubits, and show that the algorithm has a good fit with the best-known-distance codes on up to $n=20$ physical qubits as listed on the codetables.de website \cite{codetables.de}.
Calderbank-Steane-Shor (CSS) codes \cite{CSS} are an important subclass of stabiliser codes.
We apply our algorithm to find optimal distance CSS codes on up to $n=20$ physical qubits - to our knowledge the first time this analysis has been done.
We then search for optimal stabiliser codes for a biased error model where $Z$ errors occur far less frequently than $X$ and $Y$ errors. Running the algorithm for stabiliser codes encoding 1 logical qubit into 12 physical qubits, we see a significant improvement in the undetectable error rate of the resulting code versus the corresponding best-known-distance code.

\section{Background}
Stabiliser codes are one of the primary tools used in quantum error correction. 
In this work, we optimise stabiliser codes for a given Pauli error model. 
In this Section, we review background material on stabiliser codes, Pauli error models and previous work on optimisation of stabiliser codes using machine learning and other techniques. 

\subsection{Stabiliser Codes}
In this Section, we introduce the key concepts of stabiliser codes and illustrate these using the well-known 5-qubit code as an example. Stabiliser codes are defined by choosing a set $\mathbf{S}$ of \textbf{stabiliser generators.} 
The stabiliser generators are Pauli operators on $n$ qubits and generate the \textbf{stabiliser group} denoted $\braket{\mathbf{S}}$. 
The \textbf{codespace} is the subspace of states on $n$ qubits $\mathcal{H}_2^n$ fixed by all elements of $\mathbf{S}$ - that is the set $\{\ket{\psi} \in \mathcal{H}_2^n : A\ket{\psi} = \ket{\psi}, \forall A \in \mathbf{S}\}$. 
The codespace is trivial if and only if $-I$ is in the stabiliser group $\braket{\mathbf{S}}$. 
This in turn implies that all elements of $\mathbf{S}$ commute because if there exist $A,B\in\mathbf{S}$ such that $AB = -BA$, we have that:
\begin{equation}
    \ket{\psi} = AB\ket{\psi} = -BA\ket{\psi} = -\ket{\psi} \implies \ket{\psi}= 0.
\end{equation}
\subsubsection{Vector Representation of Pauli Operators}
Pauli operators on $n$ qubits have a useful length $2n$ \textbf{binary vector representation}. 
Let $\mathbf{x, z}$ be binary vectors of length $n$. 
We define an \textbf{unsigned Pauli operator} as:
\begin{equation}
X(\mathbf{x}) Z(\mathbf{z}) : = \prod_{0 \le i <n}X_i^{\mathbf{x}[i]} \prod_{0 \le i <n}Z_i^{\mathbf{z}[i]}.
\end{equation}
We refer to the vector $\mathbf{x}$ as the \textbf{X-component} and $\mathbf{z}$ as the \textbf{Z-component} of the operator.
The operators of form $X(\mathbf{x}) Z(\mathbf{z})$ generate the Pauli group on $n$ qubits, modulo global phases. 
In this paper, we will not be concerned about the phase of Pauli operators as the phase of valid stabiliser generators is defined up to a sign of $\pm 1$ and changing the sign of a stabiliser generator does not affect the error-correcting properties for the error models we consider.

Up to global phase, multiplication of Pauli operators corresponds to addition of the vector representations modulo $2$. 
This fact allows us to think of groups of Pauli operators as vector spaces over $\mathbb{F}_2$. For instance, we can calculate an \textbf{independent} set of generators of a stabiliser group by calculating the reduced row echelon form (RREF) of the \textbf{check matrix} $S$ whose rows are vector representations of the stabiliser generators. Let the rank of the matrix $S$ be $r$ and let $\braket{S}$ denote the span of $S$ over $\mathbb{F}_2$. 
The sizes of both $\braket{S}$ and the group generated by the corresponding Pauli operators $\braket{\mathbf{S}}$ are  $2^r$.

We can check whether two Pauli operators commute via the \textbf{symplectic inner product} of their vector representations\textbf{.} The Pauli operators $X(\mathbf{x}_1) Z(\mathbf{z}_1)$ and $X(\mathbf{x}_2) Z(\mathbf{z}_2)$ commute if and only if:
\begin{align}
(\mathbf{x}_1,\mathbf{z}_1) \Omega_n (\mathbf{x}_2,\mathbf{z}_2)^T = \mathbf{z}_1 \cdot \mathbf{x}_2 +\mathbf{x}_1 \cdot \mathbf{z}_2 =   0 \text{ mod } 2, \label{eq:pauli_commutation}
\end{align}
where  $\Omega_n := \begin{pmatrix}0_{n\times n}&I_n\\I_n&0_{n\times n}\end{pmatrix}$ is the \textbf{binary symplectic form}. 
We will use $\Omega$ going forward rather than $\Omega_n$ because $n$ is determined by the number of physical qubits used by the stabiliser code.

\subsubsection{Logical Pauli Operators}
Pauli operators which commute with all elements of $\mathbf{S}$ but are not in the stabiliser group $\langle \mathbf{S}\rangle$  are called \textbf{non-trivial logical Pauli operators.} 
A generating set of the non-trivial logical Pauli operators can be determined using the method of Section 10.5.7 of \cite{nielsen_chuang_2010}.
Using Gaussian elimination and permuting the qubits by some permutation $\pi$, we can write the check matrix $S$ of a stabiliser code in the following \textbf{standard form}:
\begin{align}
\begin{BMAT}(@){ccc}{cc}
\begin{BMAT}(@,22pt,12pt){c}{c}S \end{BMAT}=&
      \left(\begin{BMAT}(@,22pt,12pt){ccc|ccc}{cc}
      I&A_1&A_2&B&0&C_1\\
      0&0&0&D&I&C_2
    \end{BMAT}\right) &
      \begin{BMAT}(@,22pt,12pt)[2pt]{l}{cc}
      \rdelim\}{0.8}{1mm}[$r$]\\
      \rdelim\}{0.8}{1mm}[$s$]
    \end{BMAT}
    \\
    \begin{BMAT}(@,22pt,12pt){c}{c}\;\end{BMAT}&
     \begin{BMAT}(@,22pt,10pt){cccccc}{c}
\underbrace{}_r&\underbrace{}_s&\underbrace{}_k&\underbrace{}_r&\underbrace{}_s&\underbrace{}_k
     \end{BMAT} 
     & \begin{BMAT}(@,22pt,10pt){c}{c}
           \;
     \end{BMAT}
\end{BMAT}\label{eq:standard_form}
\end{align}
The lower block of the standard form comprises $s$ independent checks which involve strings of $Z$ operators only, whereas the upper block comprises $r$ independent checks which involve at least one $X$ operator and may also involve $Z$ operators. The rank of the check matrix $S$ determines the number of \textbf{logical qubits} $k := n - r - s$. 

We can construct a set of $k$ independent \textbf{logical Z} operators and $k$ independent \textbf{logical X} operators for the code using the method of Section 10.5.7 of \cite{nielsen_chuang_2010} as follows:
\begin{align}
L := \begin{pmatrix}L_Z\\L_X\end{pmatrix} = 
\left(\begin{array}{c c c | c c c}
      0&0&0&A_2^T&0&I\\
      0 & C_2^T&I&C_1^T&0&0
    \end{array}\right)\label{eq:logical_paulis}
\end{align}
Using the commutation relation set out in  \ref{eq:pauli_commutation}, it is easy to see that the Pauli operators defined by $L_Z$ and $L_X$ commute with the stabiliser generators. 
Each row of $L_Z$ anticommutes with the corresponding row of $L_X$, and commutes with all other rows.
The cosets $\mathbf{u}L + \braket{S}$ where $\mathbf{u}$ is a binary vector of length $2k$ correspond to classes of logical Pauli operators equivalent up to some product of stabiliser generators, and collectively form an  exhaustive set of the operators which commute with the stabiliser generators. 
All cosets apart from the trivial coset $\braket{S}$ have no overlap with the stabiliser group, and so correspond to the non-trivial logical Pauli operators.
As $S$ is of rank $n-k$ and $L$ is of rank $2k$, there are $2^{n-k}(4^k-1) = \mathcal{O}(2^{n+k})$ non-trivial logical Pauli operators in total. 

\subsubsection{Distance of a Stabiliser Code}
The \textbf{distance} $d$ of a stabiliser code is often used as a metric for the capacity of a code to detect and correct errors.
The distance is the minimum  weight of all non-trivial logical Pauli operators where the \textbf{weight} of a Pauli operator is defined as the number of qubits acted upon by a non-identity operator. 
As the number of non-trivial logical operators grows exponentially in $n + k$, finding the distance of stabiliser codes can be computationally expensive, and is in fact an NP-hard problem \cite{code_dist_NP_hard}.

\begin{exmp}[The 5-Qubit Code - Logical Operators and Distance]
\label{eg:513}
In this Example, we illustrate the key stabiliser code concepts using the well-known 5-qubit code \cite{5-qubit_code}. 
We will show that this code encodes one logical qubit and has distance $3$. 
The  stabiliser generators of the 5-qubit code are $S_0:=IXZZX$ and cyclic shifts of $S_0$ such that $S_j$ is $S_0$ shifted $j$ places to the right. 
We can also write $S_j = \pi^jS_0$ where  $\pi$ is the cyclic permutation  $\pi :=(0,1,2,3,4)$. 
In the following table, we list the stabiliser generators and their vector representations. 
Note that $S_4$ can be obtained by multiplying together stabilisers $S_0$ to $S_3$, and so is not independent - this can also be seen by adding the corresponding vector representations:
\begin{align}
\begin{array}{|c|c:c|}
\hline
\text{Stabiliser}&\mathbf{x}&\mathbf{z}\\
\hline
S_0=IXZZX&01001&00110\\
S_1=XIXZZ&10100&00011\\
S_2=ZXIXZ&01010&10001\\
S_3=ZZXIX&00101&11000\\
\hdashline
S_4=XZZXI&10010&01100\\
\hline
\end{array}
\end{align}
We put the check matrix formed from the vector representations of the stabiliser generators into the standard form of \ref{eq:standard_form} via Gaussian elimination:
\begin{align*}
S &= \left(\begin{array}{c c |c c}1000&1&1101&1\\0100&1&0011&0\\0010&1&1100&0\\0001&1&1011&1\end{array}\right) \\
&:= \left(\begin{array}{c c|c c}I&A_2&B&C_1\end{array}\right)
\end{align*}
In this case, no permutation of the qubits is required and both $A_1$ and $C_2$ are trivial. Note that $s=0$ because there are no stabiliser generators composed of only Z operators, and that $r=4$. 

As there are $n=5$ physical qubits and $4$ independent stabiliser generators, there are $k=5-4=1$ logical qubits. Applying \ref{eq:logical_paulis} and as $A_1$ and $C_2$ are trivial we find the following generating set for the logical Pauli operators:
\begin{align*}
L &= \begin{pmatrix}L_Z\\L_X\end{pmatrix} = 
\left(\begin{array}{c c | c c}
      0&0&A_2^T&I\\
      0 &I&C_1^T&0
    \end{array}\right)\\
    &= \left(\begin{array}{c|c}
      00000&11111\\
      00001& 10010
    \end{array}\right).
\end{align*}
We set the logical Z operator to be $\overline{Z}_0 := ZZZZZ$ and the logical X operator to be $\overline{X}_0 := ZIIZX$ - these operators commute with all the stabiliser generators, but anti-commute with each other. 

The weight of $\overline{X}_0$ is 3 and by exhaustively listing all $2^r(4^k-1) = 2^4(4-1) = 48$ non-trivial logical Pauli operators, we can confirm that there are no others of smaller weight. Accordingly, the distance of the 5-qubit code is $3$. As the code has 5 physical qubits, 1 encoded qubit and has distance 3 the  \textbf{parameters of the code} are  $[[ 5,1,3]]$.
\end{exmp}

\subsubsection{Canonical Form of Stabiliser Codes}
The requirement that the stabiliser generators must commute to define a non-trivial codespace means that the standard form of \ref{eq:standard_form} includes redundant information. In the following theorem, we show that any stabiliser code has a \textbf{canonical form} in terms of binary matrices which eliminates this redundant information. This allows us to write a more compact representation of stabiliser codes, and in Appendix \ref{app:encoding_operator} we show how this form also allows us to construct an \textbf{encoding operator} of particularly simple form. The canonical form is used as the genotype for our evolutionary algorithm:

\begin{thm}Any $[ [ n, k]]$ stabiliser code with $s$ independent stabilisers composed entirely of Z-operators and $r := n-k-s$ has a canonical form as follows:
\begin{enumerate}
    \item A  binary $(n-k) \times k$ matrix $C$;
    \item A symmetric $r \times r$ binary matrix $M$;
    \item A binary $r \times (n-r)$ matrix $A$;
    \item A permutation $\pi$ of the $n$ qubits.
\end{enumerate} 
\end{thm}\label{thm:canonical}
\textbf{Proof:}
We will show how to find $A, C$ and $M$ starting from the standard form of \ref{eq:standard_form}:
\begin{align*}
\begin{BMAT}(@){ccc}{cc}
\begin{BMAT}(@,22pt,12pt){c}{c}S \end{BMAT}=&
      \left(\begin{BMAT}(@,22pt,12pt){ccc|ccc}{cc}
      I&A_1&A_2&B&0&C_1\\
      0&0&0&D&I&C_2
    \end{BMAT}\right) &
      \begin{BMAT}(@,22pt,12pt)[2pt]{l}{cc}
      \rdelim\}{0.8}{1mm}[$r$]\\
      \rdelim\}{0.8}{1mm}[$s$]
    \end{BMAT}
    \\
    \begin{BMAT}(@,22pt,12pt){c}{c}\;\end{BMAT}&
     \begin{BMAT}(@,22pt,10pt){cccccc}{c}
\underbrace{}_r&\underbrace{}_s&\underbrace{}_k&\underbrace{}_r&\underbrace{}_s&\underbrace{}_k
     \end{BMAT} 
     & \begin{BMAT}(@,22pt,10pt){c}{c}
           \;
     \end{BMAT}
\end{BMAT}
\end{align*}
Set $C := \begin{pmatrix}C_1\\C_2\end{pmatrix}$ and $A:= \begin{pmatrix}A_1& A_2\end{pmatrix}$ - these have the dimensions outlined in the statement of the theorem. 

For the stabiliser generators in the first block of \ref{eq:standard_form} to commute with those in the second block, we require that:
\setstacktabbedgap{3pt}
\begin{align*}
0_{r\times s} &=\parenMatrixstack{I&A_1& A_2& |& B& 0& C_1}\Omega\parenMatrixstack{0&0&0&|&D&I&C_2}^T \\
&= \parenMatrixstack{I&A_1&A_2}\cdot\parenMatrixstack{D^T&I&C_2^T}\\ &= D^T + A_1+A_2C_2^T.
\end{align*}
Equivalently $D = A_1^T + C_2 A_2^T$ and so $D$ is completely determined once $A$ and $C$ are specified. 
We also require that the stabilisers in the first block commute with each other and so:
\begin{align*}
    0_{r \times r} &= \parenMatrixstack{I&A_1&A_2&  |&B&0&C_1}\Omega\parenMatrixstack{I&A_1&A_2&  |&B&0&C_1}^T \\
    &= \parenMatrixstack{I&A_1&A_2}\cdot\parenMatrixstack{B^T&0&C_1^T} + \parenMatrixstack{B&0&C_1}\cdot \parenMatrixstack{I&A_1^T&A_2^T}\\
    &= B^T + A_2C_1^T+ B + C_1A_2^T \\
   &= (B + C_1A_2^T) + (B + C_1A_2^T)^T.
\end{align*}
Equivalently $M:= B + C_1A_2^T$ is a binary $r \times r$ symmetric matrix, and $B$ is completely determined once $A, C$ and $M$ are specified.
Replacing $B$ and $D$, the check matrix in terms of $A, C$ and $M$ is as follows:
\begin{align}
S = 
\left(\begin{array}{c c c | c c c}
I & A_1&A_2 & M + C_1A_2^T & 0 & C_1\\
0 &0&0&A_1^T+C_2A_2^T&I&C_2
\end{array}\right)
\end{align}
\null\hfill$\Box$
\begin{exmp}[Canonical Form of 5-Qubit Code]\label{eg:5-qubit-canonical}
In this example, we show how to calculate the canonical form of the 5-qubit code. We have already calculated the matrices $A,B$ and $C$ in Example \ref{eg:513} as follows:
\begin{align*}
A = A_2 = \begin{pmatrix}1\\1\\1\\1\end{pmatrix};\;\;
B = \begin{pmatrix}1101\\
    0011\\
    1100\\
    1011\end{pmatrix};\;\;
C = C_1 = \begin{pmatrix}1\\0\\0\\1\end{pmatrix}.
\end{align*}
The symmetric matrix $M$ is calculated as follows:
\begin{align*}
    M&:=B + C_1A_2^T
    =\begin{pmatrix}0&0&1&0\\
    0&0&1&1\\
    1&1&0&0\\
    0&1&0&0\end{pmatrix}.
\end{align*}
The canonical form is the identity permutation plus the matrices $C, A$ and $M$. \null\hfill$\Box$

In Appendix \ref{app:encoding_operator}, we construct an encoding operator for the code based on the canonical form.
We show that $Q:= \begin{pmatrix}M&A\\A^T&0\end{pmatrix}$ can be interpreted as specifying a  set of $S$ operators (along the diagonal) and \textit{CZ} operators (off-diagonal entries), and that $C$ can be interpreted as specifying a set of \textit{CX} operators.
\end{exmp}

\subsubsection{Error Detection and Correction for Stabiliser Codes}
\textbf{Syndrome error correction} is used to detect and correct errors affecting a stabiliser code. This  involves measuring each of the stabiliser generators and applying a Pauli correction back into the codespace. Pauli errors $E$ which anti-commute with a stabiliser $A \in \mathbf{S}$ can be detected because $AE\ket{\psi} = -EA\ket{\psi} = -E\ket{\psi}$ and so result in an outcome of $-1$ when measuring $A$. 
Errors which are in the stabiliser group $\braket{\mathbf{S}}$ do not affect stored logical information and so can be disregarded. 
The \textbf{syndrome vector} is a binary vector of length $n-k$ and we record a value of $1$ in component $i$ if measuring the stabiliser generator $A_i$ gives an outcome of $-1$ and $0$ otherwise.
Decoders take the syndrome vector as input and find a Pauli correction back into the codespace.
Errors of weight up to $\lfloor (d-1)/2\rfloor$ can be corrected in theory due to a sphere packing argument \cite{sphere_packing}.
Finding a \textbf{minimum weight correction} is of comparable complexity to finding the distance of a code, and so is NP-hard.
Finding a \textbf{maximum likelihood} decoder which applies a correction corresponding to the most likely logical Pauli error is in fact a $\#P$-hard problem \cite{quantum_decoder_sharp_p}. 
In this work, we optimise stabiliser codes without considering whether an efficient  minimum weight or maximum likelihood decoder exists for the code.

\subsection{Pauli Error Models}
In this work, we seek to optimise stabiliser codes for a given \textbf{Pauli error model}.
We consider error models which are probability distributions over the Pauli operators on $n$ qubits. 
More formally, the error model is a map $P$ from the set of all possible unsigned Pauli operators on $n$ qubits to real numbers in the interval $[0,1]$ such that:
\begin{equation}
    \sum_{\mathbf{x, z} \in \mathbb{F}_2^n} P(X(\mathbf{x}) Z(\mathbf{z})) = 1.
 \end{equation}
In other words, error $X(\mathbf{x}) Z(\mathbf{z})$ occurs with probability $P(X(\mathbf{x}) Z(\mathbf{z}))$.
The rationale for choosing a Pauli error model is that the Pauli operators form a basis for unitary operators under complex addition. Providing we can correct  Pauli errors, we can correct any linear combination of these (this is referred to as the \textbf{discretisation of errors} see Section 10.3.1 of \cite{nielsen_chuang_2010}).
\textbf{Correlated error models} in which errors are modelled by multi-qubit gates can be described via a Pauli error model, but we do not optimise for such models in this work.  
Instead, we optimise for independent and identically distributed Pauli errors on each qubit because permuting the qubits of a stabiliser code does not affect performance for such a model. Considering equivalence classes of stabiliser codes up to permutations of the qubits greatly reduces the search space for our algorithm.

In the \textbf{depolarising error model}, Pauli errors occur independently on each qubit with probability  $p_X = p_Y = p_Z = p$ and the probability of no error occurring on the qubit is $1-3p$. 
In the depolarising model, any Pauli error of weight $d$ has probability $p^d(1-3p)^{n-d}$, and the probability of higher weight operators is exponentially lower.
Accordingly, minimising the undetectable error rate for the depolarising error model will also produce codes with maximal distance, providing we choose sufficiently small $p$, and this allows us to benchmark our algorithm versus known data for stabiliser code distances.

\textbf{Biased error models} are another commonly encountered error model for quantum devices. 
For biased error models, Pauli errors occur independently on each qubit but the probability of one type of Pauli error may be significantly lower than the other types - for instance $p_X = p_Y \gg p_Z$.  Kerr-Cat qubits are an example of such qubit architectures \cite{kerr-cat}. 
Such devices may have advantages in implementing certain quantum computing operations - examples include magic state distillation \cite{biased_noise_MSD} and state preparation \cite{biased_noise_cluster_state}. 

\subsection{Problem Definition} \label{sec:problem_definition}
In this paper we present an algorithm which takes as input a number of physical qubits $n$, a number of logical qubits $k$ and a Pauli error model. 
For all $[[n,k]]$ stabiliser codes with check matrix $S$ and logical Paulis in binary form $L$, our objective is to minimise the \textbf{undetectable error rate}:
\begin{equation}
    P_S := \sum_{A \in \braket{S,L} \setminus \braket{S}}P(A).
 \end{equation}
In the above equation, $\braket{S,L}$ means the row span of $S$ and $L$ over $\mathbb{F}_2$.
The undetectable error rate is well-defined for any Pauli error model and is an inherent property of the code independent of the choice of decoder.

Finding the $[[n,k]]$ stabiliser codes with lowest undetectable error rate is a very difficult combinatorial optimisation problem because the number of codes grows exponentially in $n$ and $k$.
The number of $[ [ n, k ] ]$ stabiliser codes is given by \cite{count_stabiliser_codes}:
\begin{align}
    \text{Stab}(n,k,2) &=  2^{n-k}\prod_{0 \le i < n-k}(4^{n-i} -1)/(2^{n-k-i} -1)\\
    &= \mathcal{O}(2^{(n-k)(n +3k+3)/2}).
\end{align}
Applying this formula, there are approximately $2^{247} \approx 2 \times 10^{74}$  possible stabiliser codes with $n=20$ and $k=1$.

\subsection{Previous Work} 
\begin{figure}[h]
    \centering
    \includegraphics[width=0.95\linewidth]{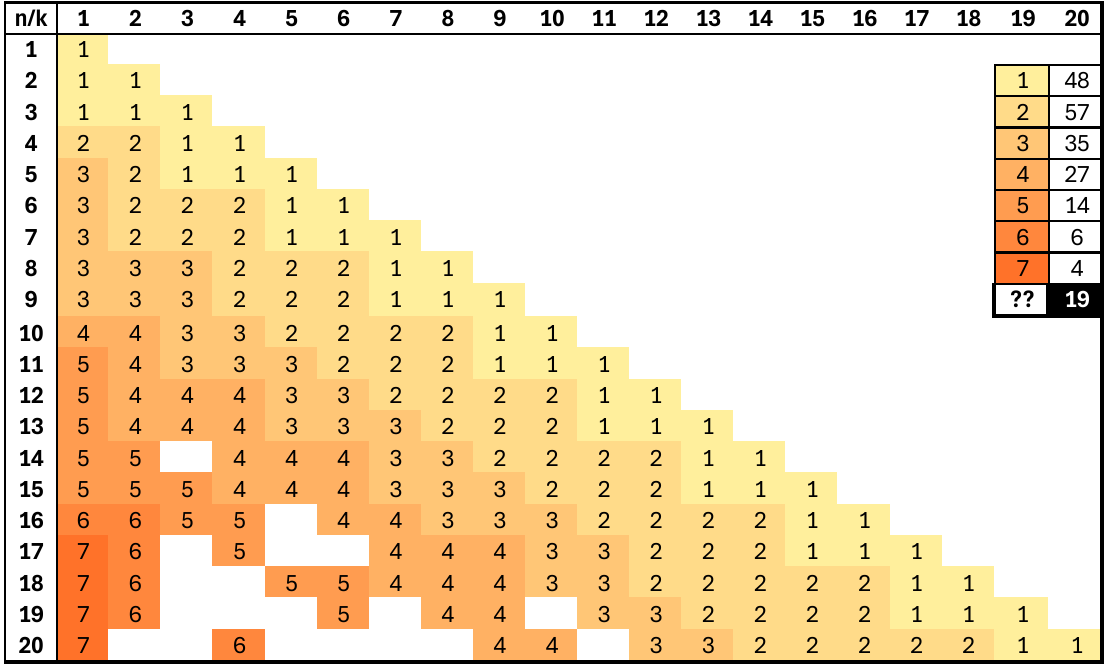}
    \caption{Best-known-distance $[[n,k]]$ stabiliser codes. The 19 blank spaces indicate combinations of $n$ and $k$ where an upper bound on distance is not known. Source: codetables.de.}
    \label{fig:best-known-codes}
\end{figure}

Optimising stabiliser codes for a depolarising error channel largely corresponds to finding the stabiliser code with the highest distance $d$ for a given number of physical qubits $n$ and logical qubits $k$.
In the late 1990s and early 2000s, there was a considerable effort to discover $[[n,k]]$ stabiliser codes which have optimal distance. 
\textbf{Linear programming bounds} for the maximum distance of $[[n,k]]$ quantum error correction codes were developed in \cite{quantum_macwilliams, quantum_shadow_enumerators, quantum_LP_bound}. 
In \cite{GF4} the authors set out construction techniques for stabiliser codes based on classical codes over $GF(4)$, and show how to construct new codes from existing codes with known distance. The authors also show how to construct \textbf{constacyclic} and \textbf{cyclic codes}, of which the 5-qubit code is an example, producing a family of perfect quantum codes with distance $3$ and optimal $n$ and $k$. A construction for \textbf{quantum BCH codes} was set out in \cite{quantum-BCH} that allows for construction of stabiliser codes with a target distance.
A table of the best-known-distance stabiliser codes for given $n$ and $k$ is available on the codetables.de website \cite{codetables.de}. 
For values of $n$ less than 20, there are a number of instances where the distance of the best possible stabiliser code is unknown (see Fig.~\ref{fig:best-known-codes}). 
Even where there is a known bound, in some cases no examples are known of codes meeting the bound. 

A different approach taken in \cite{optimisation_convex,optimisation_iterative,optimisation_cartan} is to vary the encoding and decoding operators for given $n$ and $k$ and seek to optimise the fidelity of the resulting channel in the presence of a particular error model. Optimising in this way is extremely challenging because describing a general unitary operator on $n$ qubits requires $2^{2n}$ complex numbers. 
A numerical optimisation was done in both \cite{optimisation_cartan,optimisation_iterative}, but these were limited to $n=5$ physical qubits.
In \cite{olle2024simultaneousdiscoveryquantumerror} reinforcement learning was used to search for optimal encoders for up to 15 qubits.

Optimisation of stabiliser codes has also been done via reinforcement learning in \cite{optimisation_lego} where the authors construct codes from elementary tensors via the quantum lego framework.
In \cite{optimisation_RL}, the authors use reinforcement learning to optimise for variations of the surface code on 18 physical qubits. The algorithm allows for the introduction of up to 50 new qubits which can be used to update the stabiliser generators in response to a changing error model which can include biased error models.

In terms of biased error models, the XZZX code \cite{XZZX} is an example of a stabiliser code which has good protection against errors where the error model is highly biased. 
There have been a number of studies which start with a given stabiliser code, then deform it by applying Clifford operations or changing the size of the code to improve performance. 
In \cite{tailored_robertson} the authors optimised Clifford transformations of the 7-qubit Steane code for biased error channels.
In \cite{tailored_surface} the authors assumed a biased error model and compared the performance of the square surface code to a rotated form and a form where the side-lengths of the surface code grid are co-prime. 
In \cite{kubica_surface}, the authors examine Clifford deformations of the surface code and assess their performance against various biased error models.
In \cite{tailored_XZZX} the authors show that changing the aspect ratio of the grid also improves the performance of XZZX codes in the presence of a biased error channel. 
In \cite{bias_tailored_LDPC}, the authors show how to construct a bias-tailored lifted product code which has good performance for a biased error channels.

Evolutionary strategies were used to search for optimal distance classical binary linear codes in \cite{evolutionary_classical_linear}, but to date this approach has not been tried for quantum error-correcting codes. 
Evolutionary algorithms involve searching for a global optimal solution, rather than using a particular code as a starting point. 
It can also potentially be used to search for stabiliser codes which use a large number of physical qubits.

\section{Method}
In this work, we present an evolutionary algorithm to search for an $[[n,k]]$ stabiliser code with the lowest possible  undetectable error rate for a given Pauli error model. 
In this Section, we give an overview of evolutionary algorithms and describe a genotype and fitness function for stabiliser codes suitable for use in such algorithms. 

\subsection{Overview of Evolutionary Algorithms}
Evolutionary algorithms can be used to solve global optimisation problems and fall within the category of stochastic optimisation algorithms (see \cite{Luke2013Metaheuristics} for a good introduction to such methods). 
The main features of evolutionary algorithms are as follows. 
We define a \textbf{genotype} which encodes the required information to construct an \textbf{individual} or \textbf{phenotype}.  
We also specify a \textbf{fitness function} which can be evaluated on each individual. 
The aim of the algorithm is to find the individual with optimal fitness function over all possible genotypes. 

The main steps of the algorithm are as follows. 
We specify a \textbf{population size} $\lambda$ and create an initial {population} of $\lambda$ randomly chosen individuals. 
We run the algorithm for a specified number of \textbf{generations}, or until a particular \textbf{termination condition} holds (for example, if a particular fitness target is achieved). 
We specify a \textbf{reproducing pool size} $\mu < \lambda$ of individuals with the highest fitness function which propagate to the next generation. 
We create the next generation by combining individuals by using a \textbf{crossover} method and allowing \textbf{mutation} of the resulting genotypes according to a specified probability distribution. 
We keep a record of the individual with the best fitness, and this is returned at the end of the algorithm. A summary of the evolutionary algorithm is set out in Fig.~\ref{fig:evolutionary_diag}.

\subsection{Genotype for Stabiliser Codes}

\begin{figure}[h]
    \centering
    \includegraphics[width=0.95\linewidth]{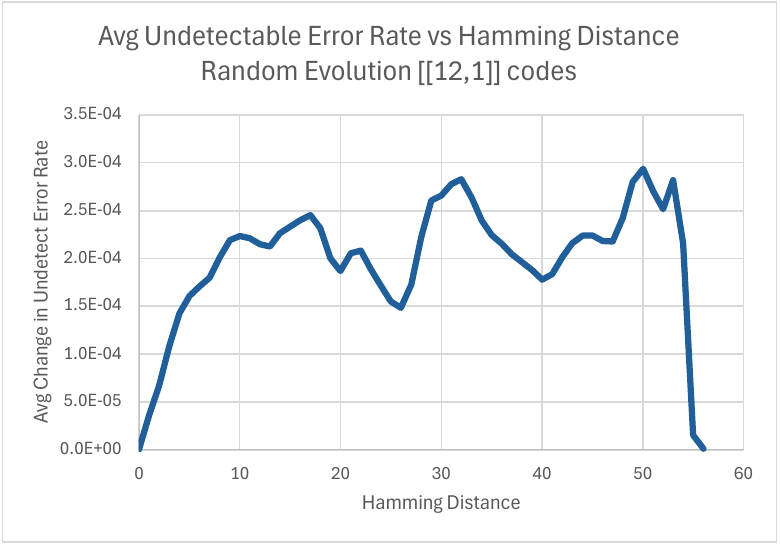}
    \caption{Hamming distance vs undetectable error rate. The data in this chart was generated by randomly choosing a binary string corresponding to a $[[12, 1]]$ stabiliser code then varying it one bit at a time to produce a population of 1000 codes. We then determined the difference between the undetectable error rate of each pair of codes and plotted this versus the Hamming distance between the codes. For small changes in Hamming distance, we see a strong correlation between Hamming distance and average change in undetectable error rate.  }
    \label{fig:hamming_dist}
\end{figure}

In this Section, we describe a genotype which can be used to encode the characteristics of an $[ [ n, k] ]$ stabiliser code. 
We use the binary matrices $A, C$ and $M$ of Thm.~ \ref{thm:canonical} to create a binary string as described in more detail below.
The length of the string depends on $n, k$ and $r$ (where $r$ is the number of independent stabiliser generators which include at least one X operator - see \ref{eq:standard_form}).
Any string of the required length corresponds to a valid $[ [ n,k]]$ code.
This allows us to generate a random initial population as well as apply crossover and random mutations for the purposes of running an evolutionary algorithm. 
Strings which are close in terms of the Hamming distance are closer than average in terms of undetectable error rate, and this is illustrated in Fig.~\ref{fig:hamming_dist}.
The following example illustrates calculating the genome of the 5-qubit code of Example \ref{eg:513}.
\begin{exmp}[Binary String Representation of 5-Qubit Code]
In this example, we represent the 5-qubit code as a binary string of length 14 based on the canonical form calculated in Ex \ref{eg:5-qubit-canonical}. 
By flattening the matrices $C$ and $A$ and taking the upper triangular half of the symmetric matrix $M$, we can represent the 5-qubit code as the following length $18$ binary string:
\begin{align}
    \left(\begin{array}{c|c|c}
        C&A&M
    \end{array}\right) :=\left(\begin{array}{c|c|c}
        1001&1111&0010,011,00,0
    \end{array}\right).
\end{align}
In Appendix \ref{app:encoding_operator}, we show that bits along the diagonal of $M$ correspond to $S$ operators in the encoding operator of the code. 
For the depolarising error model, conjugating by $S$ operators does not affect the error-correction properties of the code, so we can ignore these, resulting in a binary string representation of length $14$:
\begin{align}
    \left(\begin{array}{c|c|c}
        C&A&M
    \end{array}\right) :=\left(\begin{array}{c|c|c}
        1001&1111&010,11,0
    \end{array}\right).
\end{align}
We will in general choose a value of $r = n-k$ as this choice of $r$ maximises the possible code search space. 
The resulting binary representation for non-CSS codes uses $(n-k)(n +3k-1)/2$ bits.
We would need $4 \times 2 \times 5 =40$ bits to represent $4$ stabiliser generators on $5$ qubits, so this is a considerable saving in the number of bits to optimise. \null\hfill$\Box$
\end{exmp}

\subsection{Fitness Function}
We assess stabiliser codes against the undetectable error rate of Section \ref{sec:problem_definition}, and this is the basis of the fitness function for our algorithm. 
To calculate the undetectable error rate of an $[[n,k]]$ stabiliser code is of exponential complexity in $n+k$ because involves listing all $2^{n-k}(4^k-1) = \mathcal{O}(2^{n+k})$ non-trivial logical Pauli operators and summing the corresponding error probabilities. Accordingly, we must choose a fitness function which approximates the undetectable error rate but which can be computed relatively efficiently.

For very small codes with $n+k <= 20$ we used an exact calculation for the undetectable error rate. 
It proved possible to make an exact calculation within this regime without compromising the speed of the algorithm.
For larger values of $n+k$, an approximation method is required.

Existing distance-finding algorithms which give upper bounds on code distance (e.g. the Zimmerman algorithm or lattice methods see Sections 1.8 and 7.8 of \cite{error_correcting_linear_codes}) proved to be unsuitable for use as an approximation method. 
Firstly, in the case of biased error, the lowest weight error is not necessarily the highest probability error. 
Secondly, evolutionary algorithms perform better when optimising continuous variables rather than a single integer distance value. Stabiliser codes with the same distance may have quite different undetectable error rates, and the finer data allows the evolutionary algorithm to optimise more efficiently. Finding a lower-bound on the distance of a stabiliser code is an NP-hard problem, so we have no guarantee that it can be calculated efficiently for a given code \cite{code_dist_NP_hard}.


We used a two-stage method to approximate the undetectable error rate. 
The first stage is to search for a maximum probability generating set $L$ of $2k$ non-trivial logical Pauli generators using an evolutionary algorithm that is based on the method for binary linear codes in \cite{genetic_classical_distance} and is described in Appendix \ref{app:qDistEvol}.
The purpose of the first stage is to ensure that non-trivial logical Pauli operators with the highest probability are considered. For the depolarising channel these are the Pauli operators with lowest weight.
The second stage involves choosing an integer $t$ between $1$ and  $(n+k)/2$ and generating all linear combinations of up to $t$ rows of $S$ and $L$ as follows:
\begin{align}
    \braket{S,L}_{+t} := \left\{\mathbf{u}\begin{pmatrix}S\\L\end{pmatrix} : \mathbf{u} \in \mathbb{F}_2^{n+k}, 0 \le \text{wt}(\mathbf{u}) \le t \right\}.
\end{align}
We then calculate all linear combinations $\braket{S,L}_{-t}$ of at least  $n + k - t$ rows of $S$ and $L$ by inverting the linear combinations in $\braket{S,L}_{+t}$.
An efficient method for calculating the required linear combinations is described in Appendix \ref{app:lin_comb}.
The undetectable error rate is estimated by summing the probabilities of the non-trivial logical Pauli operators in $\braket{S,L}_{+t}$ and $\braket{S,L}_{-t}$.
Compared to calculating the undetectable error rate exactly, the approximation method is guaranteed to run in polynomial time and so is more tractable for large values of $n+k$.

\subsection{pymoo Algorithm}
The pymoo package \cite{pymoo} is a Python package with a range of  single- and multi-objective  algorithms which can be used for combinatorial optimisation tasks.
We used pymoo to develop the initial version of our evolutionary algorithm as it includes a wide range of combinatorial optimisation algorithms, crossover methods and termination conditions which allow a variety of strategies to be tested relatively quickly.

We ran the pymoo algorithm extensively to calibrate parameters for the algorithm.
We used the test case of searching for $[[12,1]]$ codes with a depolarising error model and  examined how sensitive the algorithm is to various parameters for fine-tuning. 
The upper bound on the distance of $[[12,1]]$ stabiliser codes is known to be 5, and the algorithm terminates once it finds such a code. 
The code parameters of $n=12, k=1$ have been chosen because a random search finds a distance $5$ code with very close to zero probability (a random search found a distance 5 code only once in $7.7\times 10^8$ trials). 
Because $n+k=13 \ll 20$, we can calculate the undetected error rate exactly without high computational overhead.
The algorithm parameters we varied for the pymoo algorithm were the cross type, the mutation probability and the maximum number of generations to run the algorithm.
Default parameters for the analysis are set out in Table \ref{tab:default_parameters}.
\begin{table}[h!]
\begin{center}
\begin{tabular}{|p{2.3cm} |p{0.6cm}| p{1.1cm} |p{2.8cm}| }
\hline
\textbf{Parameter} & \textbf{Value} &\textbf{Sensitivity} & \textbf{Comments}\\
\hline
Maximum Number of Generations & 1000 & High & Combines fast run-time and high accuracy\\
\hline
Cross Type & None & High &Better than any cross method\\
\hline
Mutation Rate & 0.05 & Moderate & Local maximum\\
\hline
Qubit Error Rate  & 0.01 & Low & Local maximum\\
\hline
\end{tabular}
\caption{Default parameter settings for pymoo evolutionary algorithm}
\label{tab:default_parameters}
\end{center}
\end{table}
 
\subsubsection{Sensitivity - Maximum Number of Generations}
\begin{figure}[h]
    \centering
    \includegraphics[width=0.95\linewidth]{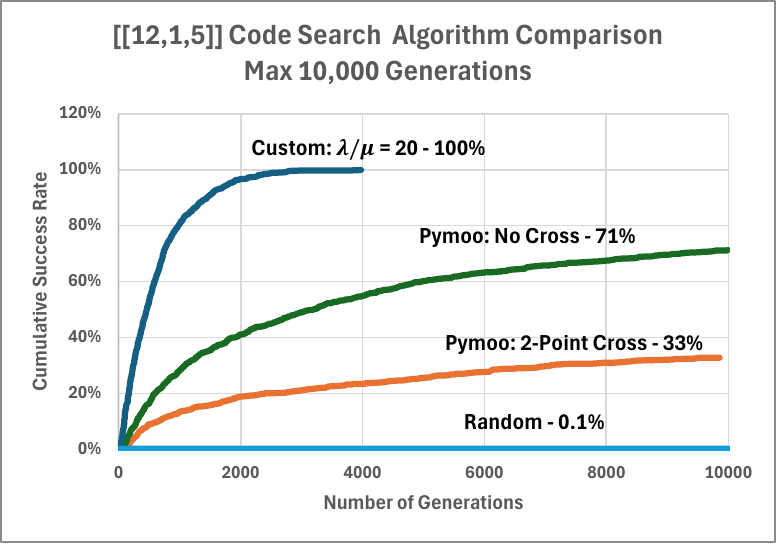}
    \caption{Search algorithm comparison - cumulative success rate of finding a $[[12,1,5]]$ stabiliser code vs number of generations, depolarising error model. The pymoo genetic algorithm using no-cross had a success rate of 71\% over 10 000 generations and 1000 trials, versus the 2-point cross with a success rate of 33\%. 
    This is a strong indication that crossover is detrimental to the search algorithm.
    The custom algorithm with no crossover and a $\lambda/\mu$ ratio of $20$ found a distance 5 code in all cases within 4,000 generations. 
    Random generation of the same number of $[[12,1]]$ codes found a code of distance 5 once over $7.7 \times 10^8$ trials, giving a strong indication that the evolutionary algorithm gives better results than random generation of codes.}
    \label{fig:sensitivity-maxgen}
\end{figure}

The accuracy of the algorithm improves when we increase the maximum number of allowed generations. In Fig.~\ref{fig:sensitivity-maxgen} we examine the cumulative probability of finding a $[[12, 1, 5]]$ code after 10 000 generations for various algorithms. 
We see that even after a relatively few number generations, it is clear which algorithm gives better results at the 10 000 generation mark.
We set a default maximum generation value of 1000 generations which combines fast run-time with a good accuracy level.

\subsubsection{Sensitivity - Cross Type}
\begin{figure}
    \centering
    \includegraphics[width=0.95 \linewidth]{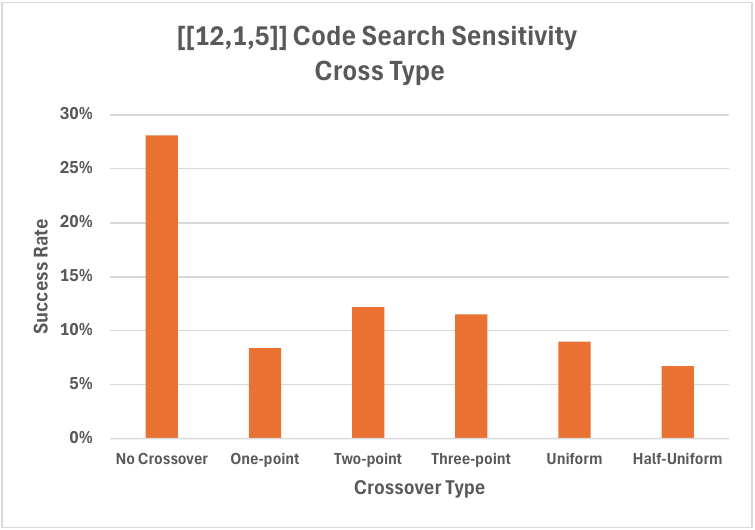}
    \caption{Sensitivity of pymoo algorithm to cross type - $[[12,1]]$ stabiliser codes, depolarising error model.}
    \label{fig:sensitivity-cross}
\end{figure}

Evolutionary algorithms typically involve selecting the individuals with the best performance in each generation then crossing them to create the next generation of individuals. 
We explored the following cross types:
\begin{itemize}
    \item \textbf{No Crossover}: the next generation consists of mutations of the parents, with no crossover
    \item \textbf{1-Point Crossover}: choose a location in the binary string representation and swap the first sections of the parents to create new individuals;
    \item \textbf{2-Point Crossover}: choose two locations in the binary string representation and swap the middle sections of the parents to create new individuals;
    \item \textbf{3-Point Crossover}: as above, but choose three locations;
    \item \textbf{Uniform Crossover}: choose each bit at random from each parent;
    \item \textbf{Half-Uniform Crossover}: for bits which differ between the parents, choose bits at random from each parent.
\end{itemize}
Running the pymoo algorithm with default parameters 1000 times we established that not using a cross function gives the highest success rate in finding a  $[[12,1,5]]$ code, with the 2-point cross next best (see Fig.~\ref{fig:sensitivity-cross}) and this is consistent with the findings for binary linear codes in \cite{evolutionary_classical_linear}.

\subsubsection{Sensitivity - Mutation Rate}
\begin{figure}
    \centering
    \includegraphics[width=0.95 \linewidth]{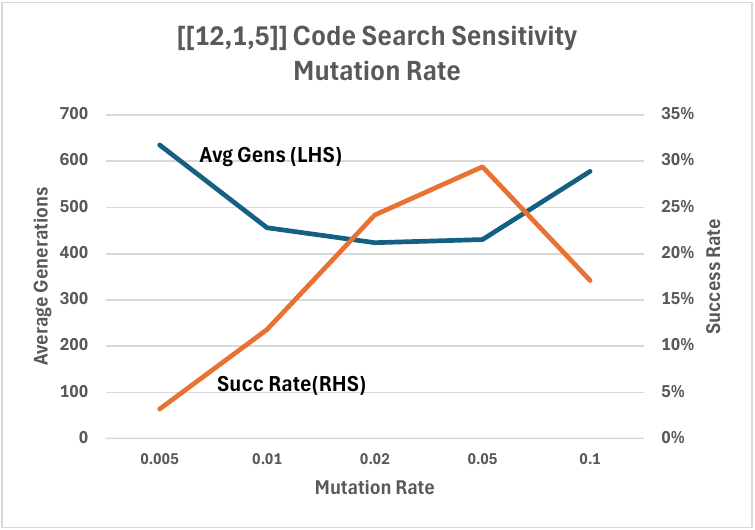}
    \caption{Sensitivity of pymoo algorithm to mutation rate - $[[12,1]]$ stabiliser codes, depolarising error model. Average number of generations on left axis and success rate on the right axis.}
    \label{fig:sensitivity-pMut}
\end{figure}

The evolutionary algorithm is moderately sensitive to the mutation rate. A sensitivity analysis suggests that a relatively low mutation probability of $0.05$ is optimal (see Fig.~\ref{fig:sensitivity-pMut}).

\subsection{Custom Algorithm}\label{sec:custom_algorithm}
Based on the insights from the pymoo algorithm, we then wrote a custom algorithm which runs faster and allows for further parameter optimisation (notably the $\lambda/\mu$ ratio of total population size to reproducing pool size).
\subsubsection{Custom Algorithm Description}
The custom algorithm has no crossover and a low mutation rate, as suggested by the results from the pymoo algorithm. 
The next generation is produced by selecting the best $\mu$ individuals, and producing $\lambda/\mu$ mutations of each.
Mutation is performed by flipping a single bit in the binary code representation.
The custom algorithm has a population size $\lambda$ equal to the number of bits in the binary representation of the $[[n,k]]$ code ($77$ bits in the case of $[[12,1]]$ codes).

\subsubsection{Sensitivity - Reproducing Pool Size Ratio}
\begin{figure}
    \centering
    \includegraphics[width=0.95 \linewidth]{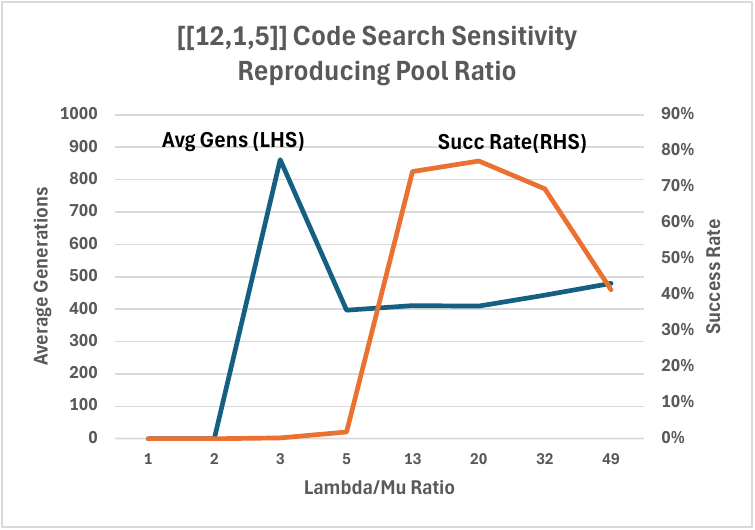}
    \caption{Sensitivity of Custom algorithm to reproducing pool size ratio - $[[12,1]]$ stabiliser codes, depolarising error model, custom algorithm. Average number of generations on left axis and success rate on the right axis.}
    \label{fig:sensitivity-lambmu}
\end{figure}

The relative size of the reproducing pool to the total population is an important parameter in evolutionary algorithms.
A higher $\lambda/\mu$ ratio means that a given individual has a higher number of offspring in the next generation, and so favours propagation of successful individuals versus a broad exploration of the search space.
Because $\lambda$ can vary widely depending on $n$ and $k$, we chose to vary $\mu$ by choosing values of $\log(\lambda/\mu)$ between $0$ and $0.9$ resulting in $\lambda/\mu$ ratios between $1$ and $49$. 
We found that the algorithm was highly sensitive to this parameter and the value $\log(\lambda/\mu) = 0.3$ (corresponding to $\lambda/\mu = 20$) to be optimal for the $[[12,1]]$ code search (see Fig.~\ref{fig:sensitivity-lambmu}).

\section{Results}
In this Section, we discuss the results of running the Custom evolutionary algorithm of Section \ref{sec:custom_algorithm}. 
We first search for optimal distance $[[n,k]]$ codes for $1 \le k < n \le 20$ by using a depolarising error model. The purpose of this exercise is to benchmark the algorithm versus the best known codes of \cite{codetables.de}. We then set out a genotype for CSS codes (i.e. stabiliser codes where each stabiliser generator can be written as either a string of $X$ operators or a string of $Z$ operators) and search for optimal codes in the range $1 \le k < n \le 20$ with a depolarising error model. We compare the difference between the difference in distance between the resulting CSS codes and the best-known-distance stabiliser codes. Finally we search for optimal $[[12,1]]$ codes for a biased error model where $p_X = p_Y \gg p_Z$.

\subsection{Distance Benchmark Versus Best-Known-Distance Codes}\label{sec:best-known}
\begin{figure}
    \centering
    \includegraphics[width=0.95 \linewidth]{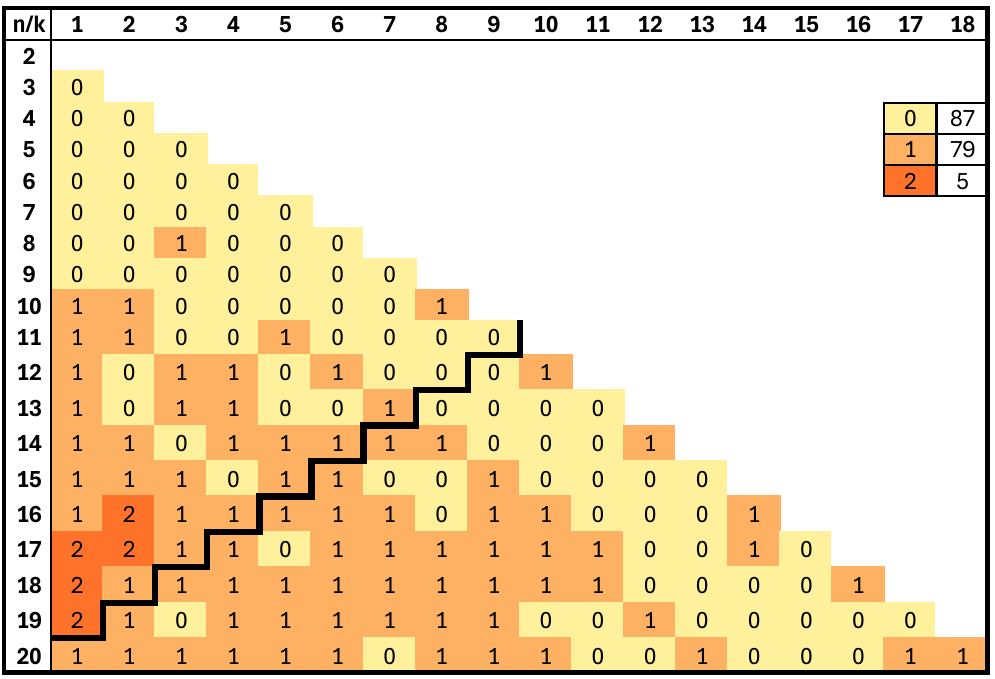}
    \caption{Difference between distance of best-known-distance-code and random search result, $1 \le k < n \le 20$, depolarising error model, average over 10 000 runs.}
    \label{fig:nRange-random}
\end{figure}

\begin{figure}
    \centering
    \includegraphics[width=0.95 \linewidth]{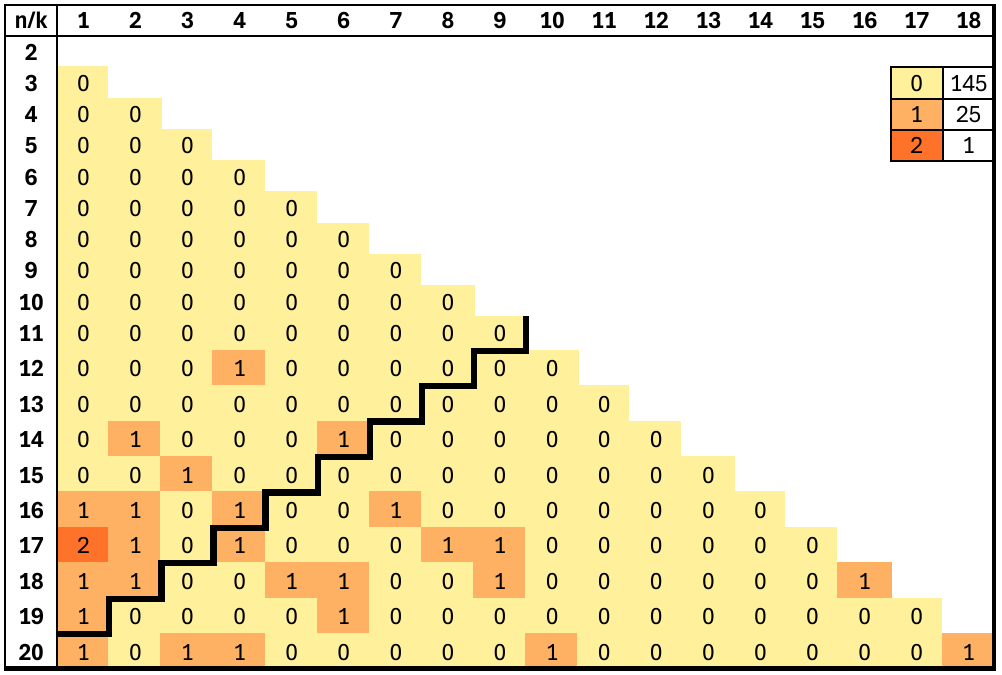}
    \caption{Difference between distance of best-known-distance-code and search result, Custom algorithm, $1 \le k < n \le 20$, depolarising error model, 10 runs, 1000 maximum generations. Black zig-zag line separates exact calculation of fitness function $(n+k \le 20)$ with approximation. The custom algorithm found a code with best-known-distance in 85\% of $[[n,k]]$ combinations.}
    \label{fig:nRange-delta-d}
\end{figure}

We conducted a search for stabiliser codes of the best known distance for the range $1 \le k < n \le 20$ with a depolarising error model. 
We set the maximum number of generations to be 1000, and conducted $10$ runs of the algorithm to assess the average success rate and average number of generations taken. The results are displayed in Fig.~\ref{fig:nRange-delta-d}. For reference, we compare these to the average success rate for randomly generated codes in Fig.~\ref{fig:nRange-random}. 

The custom algorithm successfully found a code with the same distance as the best-known-distance code in 145 of the 171 possible $[[n,k]]$ combinations, a success rate of 85\%. 
In only one case did the algorithm fail to find a code within one of the best-known-distance. 

The purpose of this analysis is to establish that the algorithm gives reasonable results versus known data over a range of code parameters.
We note that the maximum number of generations was relatively small at 1000, and that only 10 runs were performed. 
If we were to perform a deeper optimisation for a particular number of physical and logical qubits, we could use a much larger number of generations and multiple runs. 
We could also factor previous knowledge about constructions of codes with good parameters (for example those listed in \cite{GF4}) into the initial population rather than making a random population, and this is a possible area for further improvement.

\subsection{Restricting Search to CSS Codes}\label{sec:CSS}
Having calibrated the algorithm versus the data for the best-known-distance stabiliser codes, we now turn to finding optimal CSS (Calderbank-Shor-Steane) codes \cite{CSS}. CSS codes are an important subclass of stabiliser codes where each stabiliser generator can be written as either a string of $X$ operators (X-checks) or a string of $Z$ operators (Z-checks). Examples of such codes include the toric code \cite{toric} and the colour code \cite{colour_code}. To our knowledge, there has been no previous work on finding optimal distance CSS codes over this range of values for $n$ and $k$. 

To search for CSS codes, we use a modified genotype which is as follows. Let $S$ be the check matrix of a CSS code and let $r$ be the number of independent X-checks and $s$ the number of independent Z-checks such that $n = r + s + k$. Using the canonical form of Thm.~ \ref{thm:canonical} but setting $C_1$ and $M$ to zero, we can write the check matrix of a CSS code in the following form:
\begin{align}
    S &= \left(\begin{array}{c c c | c c c}
    I & A_1 &A_2 &0&0&0\\
    0&0&0&A_1^T + C_2A_2^T&I&C_2
    \end{array}\right).
\end{align}
The binary matrices  $A_1,A_2$ and $C_2$ have dimensions $r \times s, r \times k$ and $s \times k$ respectively. 
As the entries $A_1^T + C_2A_2^T$ are fully determined by $A_1,A_2$ and $C_2$, CSS codes can be represented as binary vectors of length $k(n-k) + rs$.

\begin{figure}
    \centering
    \includegraphics[width=0.95 \linewidth]{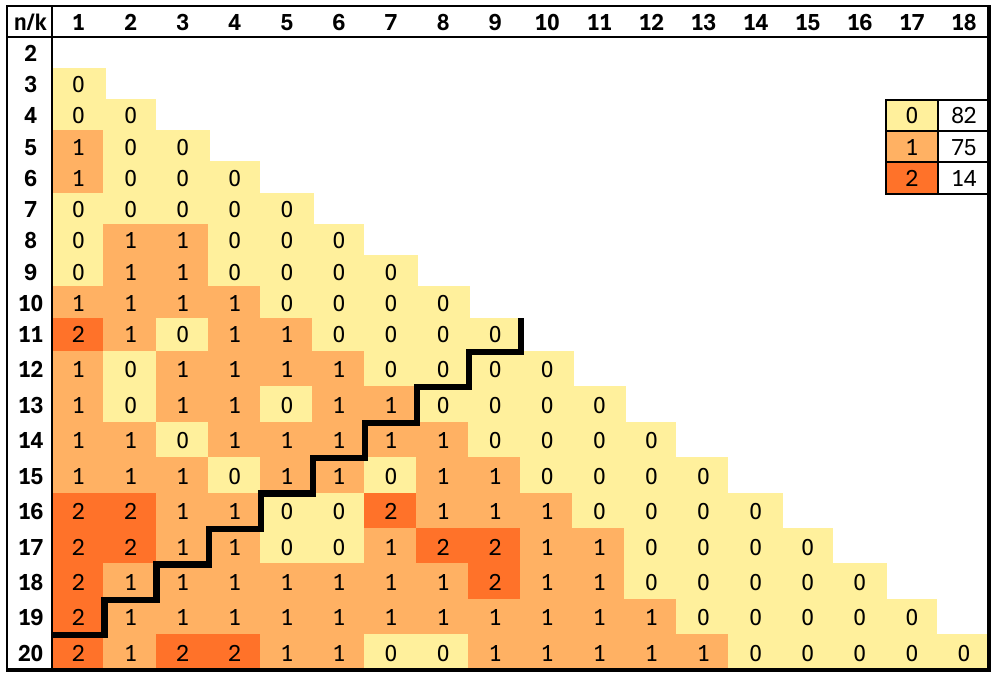}
    \caption{Difference between distance of best-known-distance code and highest distance CSS code found using evolutionary algorithm, $1 \le k < n \le 20$, depolarising error model, 2 runs, 1000 maximum generations. Black zig-zag line separates exact calculation of fitness function $(n+k \le 20)$ with approximation.}
    \label{fig:nRange-CSS}
\end{figure}

In Fig.~\ref{fig:nRange-CSS} we compare the distance of the best CSS code found using the evolutionary algorithm to the best-known-distance stabiliser code listed on \cite{codetables.de} for $1 \le k < n \le 20$. For this analysis, we chose $r := \lfloor (n-k)/2\rfloor$. For many code parameters, there exists a CSS code with the same distance as the best-known-distance stabiliser code, and for the majority of code parameters the difference is only $1$. The maximum difference in distance is $2$, and this occurs in only 14 cases.


\subsection{Biased Error Model}\label{sec:biased-search}
\begin{figure}[h]
    \centering
    \includegraphics[width=0.95 \linewidth]{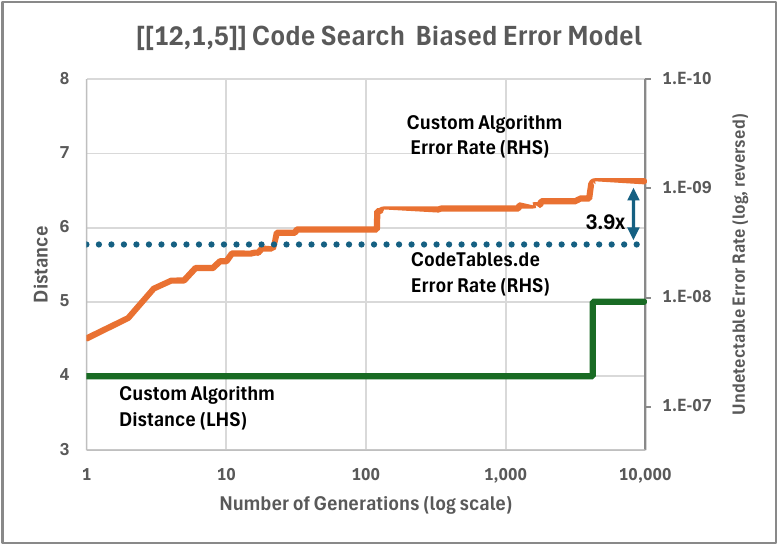}
    \caption{Optimisation of $[[12,1]]$ codes, biased error model with $p_X=p_Y = 0.01; p_Z = 0.001$, Custom algorithm. Single run with undetectable error rate and distance noted for each generation, and comparison to base $[[12,1,5]]$ code listed on codetables.de. Distance is on the left axis and undetectable error rate is on the right axis. }
    \label{fig:biased-12-1-search}
\end{figure}

We next applied the custom algorithm to search for $[[12,1]]$ codes with a biased error model where the likelihood of a Z error is much lower than an X or Y error.
We chose the following probabilities of an X, Y or Z error occurring independently on each qubit: $p_X=p_Y = 0.01; p_Z = 0.001$.
A single run of the algorithm was done with a maximum of 10 000 generations. 
We chose to allow for $S$ operators along the symmetric matrix $M$ of Thm.~ \ref{thm:canonical} and chose $r = n-k$.
The undetectable error rate of the best code found by the algorithm at each generation was compared to a base code - the $[[12,1,5]]$ best-known-distance code from codetables.de - and we also tracked the distance of the best code at each generation.
The custom algorithm found codes with a lower undetectable error rate than the base code after 22 generations.
After 4,224 generations, the algorithm found a distance 5 code, which is the highest possible distance for a $[[12,1]]$ code. 
The best code found had an undetectable error rate of $8.54\times 10^{-10}$ which was $3.9$ times lower than the base code (see Fig.~\ref{fig:biased-12-1-search}).  

\section{Discussion}
The custom algorithm of Section \ref{sec:custom_algorithm}, though quite simple, produced good results. 
Recall that the custom algorithm does not involve any crossing to produce the next generation.  
Mutation is done by changing a single bit in the string representation of the genome.
A relatively high $\lambda/\mu$ ratio appears to be optimal, meaning that the algorithm focuses on producing a large number of offspring via mutation for a small number of parents. 

Using the custom algorithm, we can identify stabiliser codes at the best known distance for almost all $n$ and $k$ less than 20 (Section \ref{sec:best-known}). 
We were able to identify CSS codes with optimal distance over the same range, and found that in almost all cases the difference in distance was at most 1 compared to the best-known-distance stabiliser codes (Section \ref{sec:CSS}). 
Finally, we showed that the undetectable error rate of the $[[12,1]]$ code can be improved by a factor of almost 4 times compared to the base $[[12,1,5]]$ best-known-distance code listed on codetables.de (Section \ref{sec:biased-search}).

The main limitation of the evolutionary algorithm is that exact calculation of the fitness function is of exponential complexity in $n+k$. 
We have addressed this by using an approximation method which finds a high probability generating set of non-trivial logical Paulis via an evolutionary algorithm. 
Even though the accuracy of the approximation deteriorates for large $n+k$, the algorithm still found optimal distance codes for most values of $1 \le k < n \le 20$, even where $n+k > 20$.
It would also be interesting to compare other optimisation methods  to our evolutionary algorithm. For instance, our algorithm produces many examples of codes with good performance, and this data could as training data for a neural network or reinforcement learning system.

We have also restricted our method to optimising for Pauli error models. We have not allowed for correlated or non-Pauli error models, though this could be a possible future research direction. 

Finally, we have sought to optimise codes based on the undetectable error rate and have not considered whether the codes also have an efficient decoder. Finding an efficient decoder is a very difficult problem, but perhaps an evolutionary algorithm might also be useful in this context.

\section{Conclusion}
Our results suggest that evolutionary algorithms have a useful role in tailoring stabiliser codes for particular devices and Pauli error models. The evidence we have presented for this is as follows. Firstly, using an evolutionary algorithm to search for stabiliser codes with the best known distance for given parameters $n$ and $k$ results in significantly higher probability of success than using randomly selected stabiliser codes. 
Secondly, the method can be used to search for codes in parameter spaces  where optimal  codes are not known - we have demonstrated this functionality in searching for CSS codes and biased error models. Thirdly, we have demonstrated that the algorithm results in codes which have a significantly lower undetectable error rate for biased error models versus the best distance $[[12,1]]$ code. This suggests that using bias-tailored codes for devices with a biased error channel can lead to significant gains.

The algorithms as implemented start with a random selection of stabiliser codes. 
We expect that starting instead with codes constructed using the methods in \cite{GF4} and which are known to have good distance would lead to significantly better results.

It would be of great interest to optimise stabiliser codes for a particular device using the methods outlined. This would include incorporating the error model and connectivity constraints of the device.

\section*{Acknowledgement}
The authors thank our ROARQ collaborators Lajos Hanzo, Balint Koczor and Zhenyu Cai for useful discussions and feedback on drafts of this paper.
MW thanks S. Shukla, G. Umbrarescu, O. Higgott, N. Shutty and Prof. T. Cubitt for helpful conversations during preparation of this work.
The authors acknowledge the use of the UCL Kathleen High Performance Computing Facility (Kathleen@UCL), and associated support services, in the completion of this work.
\bibliography{bibliography}
\bibliographystyle{IEEEtran}

\appendix 

\subsection{Canonical Encoding Circuits of Stabiliser Codes}\label{app:encoding_operator} 
In this Appendix, we show how to use the canonical form of stabiliser codes of Thm.~ \ref{thm:canonical} to construct a particularly simple encoding circuit for the code.
Quantum information is encoded into a stabiliser code by using a Clifford operation called an \textbf{encoding operator}.
We now show how to construct an encoding operator for a stabiliser code given in terms of a check matrix $S$ and the logical Paulis $L$ obtained by the method in \ref{eq:logical_paulis}. 
We first construct a set of \textbf{destabiliser generators} $R$ which have the property that each row of $R$ anticommutes with the corresponding row of $S$, but commutes with all other rows of $S$, $L_Z$ and $L_X$:
\begin{align}
    R&:=\begin{pmatrix}R_X\\R_Z\end{pmatrix} = \left(\begin{array}{c c c | c c c}
      0&0&0&I&0&0\\
      0&I&0&0&0&0
    \end{array}\right)
\end{align}
We then write the stabilisers, destabilisers and logical Paulis in \textbf{tableau format} \cite{simulation} as follows:
\begin{align}
\tau := 
\begin{pmatrix}S_X\\S_Z\\L_Z\\R_X\\R_Z\\L_X\end{pmatrix} = 
\left(\begin{array}{c c c | c c c}
I & A_1&A_2 & B & 0 & C_1\\
0 &0&0&D&I&C_2\\
0&0&0&A_2^T&0&I\\
\hline
0&0&0&I&0&0\\
0&I&0&0&0&0\\
0&C_2^T&I&C_1^T&0&0
\end{array}\right)
\end{align}
By construction, $\tau$ is a \textbf{binary symplectic matrix} because the commutation relations of $S,L_Z,R$ and $L_X$ are equivalent to the identity $\tau\Omega \tau^T = \Omega$ where $\Omega$ is the binary symplectic form of \ref{eq:pauli_commutation}. 
The Clifford operator $\mathcal{C}:\mathcal{H}_2^n \rightarrow \mathcal{H}_2^n$ corresponding to the binary symplectic matrix $\tau$ maps the $k$-qubit \textbf{logical state} $\ket{\psi}$ to the $n$-qubit \textbf{encoded state} as follows:
\begin{align}
\overline{\ket{\psi}} := \mathcal{C}\Big(\ket{+}^{\otimes (r + s)}\ket{\psi}\Big).
\end{align}



The canonical form of stabiliser codes of Theorem \ref{thm:canonical} allows us to write an encoding operator for the code in a  particularly simple form as outlined in the following Corollary: 
\begin{cor}
Given $C, A, M, \pi, n, r$ and $k$ as in Theorem \ref{thm:canonical} and $Q:=\begin{pmatrix}M & A\\A^T & \mathit{0}\end{pmatrix}$, the encoding circuit of the stabiliser code is given by 
\begin{align}
\mathcal{C} := \pi \Big(\prod_{r \le i <n} \text{Had}_i\Big) \textit{CZ}_Q \textit{CX}_C,
\end{align}
where:
\begin{align}
\textit{CZ}_Q &:=\prod_{0\le i < n}S_{i}^{Q_{ii}}  \prod_{0\le i < j < n}  \textit{CZ}_{ij}^{Q_{ij}} \\
\textit{CX}_C &:= \prod_{\substack{0\le i < n-k\\ 0 \le j < k}} \textit{CX}_{i,n-k+j}^{C_{ij}}.
\end{align}
\end{cor}
\textbf{Proof: }
The stabiliser, logical Pauli and destabiliser generators of $S$ in tableau format  are as follows:
\begin{align}
\tau = 
\left(\begin{array}{c c c | c c c}
I & A_1&A_2 & M + C_1A_2^T & 0 & C_1\\
0 &0&0&A_1^T + C_2A_2^T&I&C_2\\
0&0&0&A_2^T&0&I\\
\hline
0&0&0&I&0&0\\
0&I&0&0&0&0\\
0&C_2^T&I&C_1^T&0&0
\end{array}\right).
\end{align}
Applying Hadamard operators to the last $n-r$ qubits, we obtain:
\begin{align}
U^H := 
\left(\begin{array}{c c c | c c c}
I & 0 & C_1 & M + C_1A_2^T & A_1&A_2\\
0&I&C_2&A_1^T + C_2A_2^T   &0&0\\
0& 0&I&A_2^T&0&0\\
\hline
0&0&0&I&0&0\\
0&0&0&0& I&0\\
0&0&0&C_1^T& C_2^T&I
\end{array}\right)
\end{align}
Define the following symplectic matrices:
\begin{align}
CX_C &:=  
\left(\begin{array}{c c c | c c c}
I & 0 & C_1 & 0 & 0&0\\
0&I&C_2&0  &0&0\\
0& 0&I&0&0&0\\
\hline
0&0&0&I&0&0\\
0&0&0&0& I&0\\
0&0&0&C_1^T& C_2^T&I
\end{array}\right),\\
CZ_Q&:= 
\left(\begin{array}{c c c | c c c}
I & 0 & 0 & M  & A_1&A_2\\ 
0&I&0&A_1^T   &0&0\\
0& 0&I&A_2^T&0&0\\
\hline
0&0&0&I&0&0\\
0&0&0&0& I&0\\
0&0&0&0& 0&I
\end{array}\right).
\end{align}
It is easy to verify that  $CX_C CZ_Q  = U^H$ and that $CX_C$ and $CZ_Q$ correspond to the Clifford operations in the statement of the Corollary.\null\hfill$\Box$


\subsection{Approximation of Undetectable Error Rate}
Calculating the undetectable error rate is of exponential complexity in $n+k$, for for larger values of $n+k$ we must use an approximation method.

Our approximation method involves two stages. We first calculate a set of maximum probability generators $L$ for the logical Pauli group using an evolutionary algorithm. The second stage involves forming all linear combinations involving up to $t$ of the stabiliser generators $S$ and logical generators $L$, then calculating the total probability of the combinations.

\subsubsection{Evolutionary Algorithm for Maximum Probability Errors}\label{app:qDistEvol}
In \cite{genetic_classical_distance} an evolutionary algorithm is described for finding the distance of classical linear codes.
The \textbf{genotype} for the algorithm is a \textbf{permutation} of the $n$ columns of the generator matrix of the linear code.
Low-weight codewords are calculated by permuting the columns of the generator matrix and finding the reduced row echelon form.
The authors show that by taking an initial population of random permutations then running an evolutionary algorithm they are able to find a column permutation which produces a row whose weight is the code distance.

In \cite{qDistRnd}, this method was extended to quantum error correction codes resulting in the QDistRnd package. 
In the case of QDistRnd, random permutations are used rather than an evolutionary algorithm to find a minimum weight non-trivial logical Pauli operator.

The algorithm used in this work QDistEvol 
 uses an \textbf{evolutionary algorithm} to find high probability undetectable errors. 
Instead of searching for low-weight non-trivial logical Paulis, the algorithm searches for \textbf{high probability} non-trivial logical Paulis based on an error model given as input.
In particular, this allows the algorithm to be used for biased error models.

The QDistEvol algorithm searches for a high probability \textbf{generating set of logical Paulis}, rather than a single low-weight logical Pauli. 
This makes the algorithm suitable for generating linear combinations of high probability errors in combination with the stabiliser generators.
Rather than using random permutations, a population of permutations is maintained for selection and mutation according to an evolutionary algorithm. 
At each generation, the permutations which give the largest increase in the total probability of the generating set of logical Paulis are selected.
Mutation involves adding a \textbf{single additional transposition} to the permutation 

The algorithm appears to work well across a wide range of code types and benchmarking versus other distance-finding algorithms is likely to be an area of future work.

\subsubsection{Generating Linear Combinations of a Binary Matrix}\label{app:lin_comb}
In this Section, we demonstrate an efficient algorithm for generating linear combinations of a binary matrix. 
The algorithm takes as input an $r \times n$ binary matrix $A$ and a maximum depth $t$.
It returns all linear combinations modulo 2 involving up to $t$ rows of $A$.

The  computational cost of the algorithm is governed by the number of additions modulo 2 that are performed, and the algorithm has been designed to minimise these.
The algorithm is recursive and is based on the following binomial identity for $r,t \ge 1$:
\begin{align}
    \binom{r}{t} = \binom{r-1}{t-1} + \binom{r-1}{t}.\label{eq:bimom_id}
\end{align}
The above identity can be interpreted in the following way.
The left hand side can be thought of as the number of linear combinations of exactly $t$ of the $r$ rows of $A$.
The right hand side can be interpreted as counting the number of linear combinations of $t$ rows of $A$ which include  the final row of $A$ plus the number of linear combinations which do not include the final row of $A$. 
All linear combinations which do include the final row can be generated by finding all linear combinations of $t-1$ of the first $r-1$ rows of $A$, then adding the final row of $A$ to each linear combination. Linear combinations which do not include the final row can be found by generating all linear combinations of $t$ of the first $r-1$ rows of $A$.
Our method involves the following steps. We first allocate space for a series of binary matrices $B_v$ for $0 \le v \le t$ of size $\binom{r}{v} \times n$ which represent combinations of exactly $v$ of the $r$ rows of $A$. We set $B_0$ to be the all zero vector and $B_1 := A$. For $v \ge 2$, the algorithm populates the first $\binom{r-1}{v}$ rows of $B_v$ via a recursive call. 
It then populates the rest of $B_v$ by adding the final row of $A$ to the first $\binom{r-1}{v-1}$ rows of $B_{v-1}$.
This method requires $\sum_{1 < v \le t}\binom{r}{v}$ addition operations because for $v > 1$ we only need a single addition to generate a row of $B_v$ from previously calculated entries.
A naive method of generating all linear combinations of $v$ of the $r$ rows of $A$ requires $v-1$ additions for each of the $\binom{r}{v}$ such combinations. Hence, to generate all combinations of up to $t$ rows of $A$ would take the following number of operations:
\begin{align}
    \sum_{1 < v \le t} (v-1)\binom{r}{v} \gg \sum_{1 < v \le t}\binom{r}{v}.
\end{align}

\begin{IEEEbiography}
[{\includegraphics[width=1in,height=1.25in,clip,keepaspectratio]{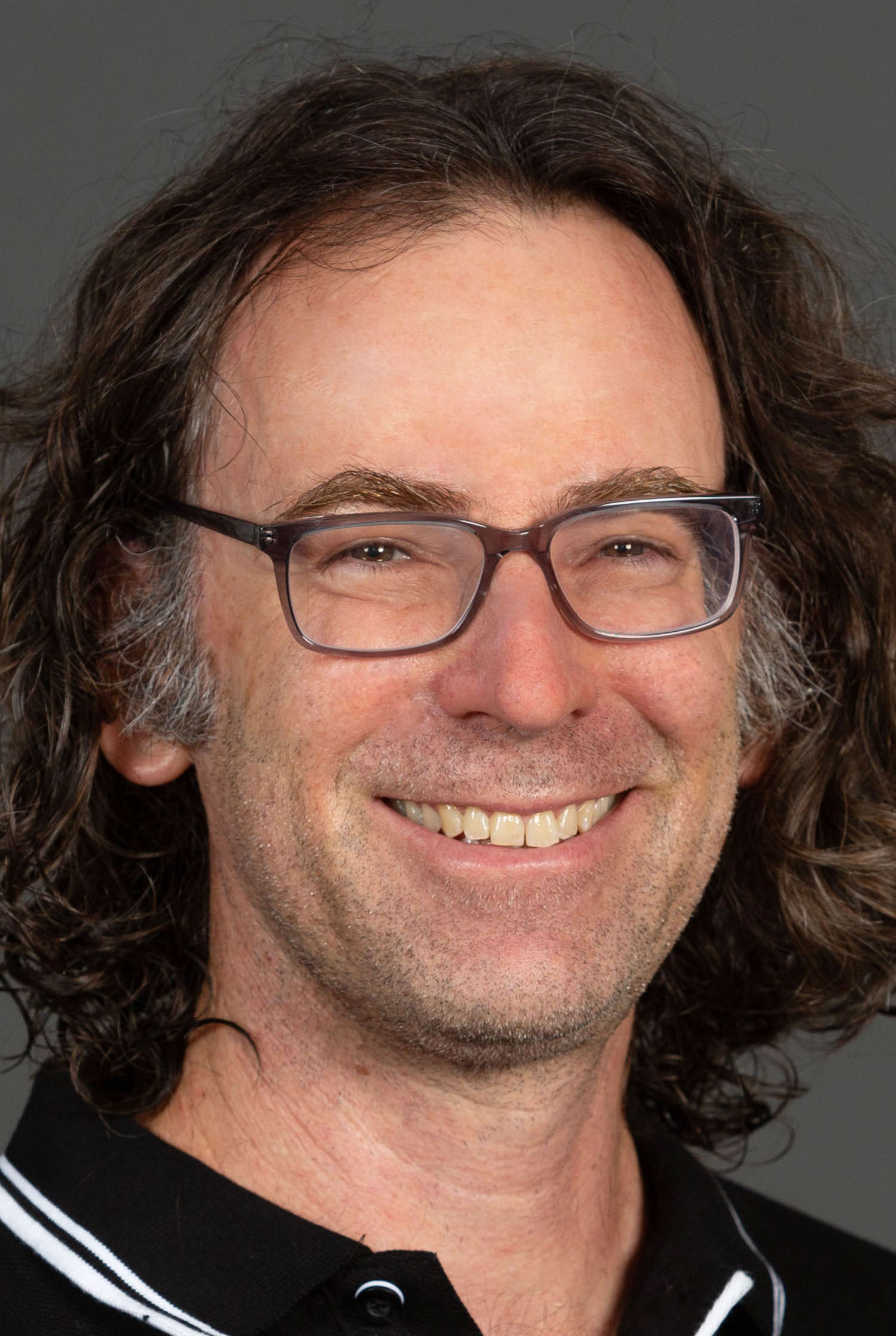}}]
{Mark Webster} completed a Bachelor of Laws (Hons) and Bachelor of Science (Hons, University Medal) combined degree from the University of Sydney.
He received a PhD in quantum computing from the University of Sydney in 2024. 

He is currently a Research Fellow at University College London (UCL). 

His research is in the quantum error correction field.
\end{IEEEbiography}

\begin{IEEEbiography}
[{\includegraphics[width=1in,height=1.25in,clip,keepaspectratio]{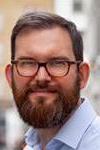}}]
{Prof. Dan Browne} received M.Sci. and Ph.D. degrees from the Imperial College London in 2000 and 2004, respectively. 

He is currently a Professor in Physics at University College London (UCL). He was a Junior Research Fellow with the Merton College Oxford, from 2004 to 2007, and was appointed as a Lecturer at UCL in 2007. He was promoted to a Reader in 2013 and to a Professor in 2017. 

He is a member of the Institute of Physics. His research is on the theory of quantum computing, including fault-tolerant quantum computation, architectures, and applications of near-term quantum computers.
\end{IEEEbiography}

\EOD

\end{document}